\documentclass[twocolumn]{svjour3} 
\PassOptionsToPackage{hyphens}{url}
\usepackage{amsmath}
\usepackage{lineno}
\usepackage{graphicx}
\usepackage[round,authoryear]{natbib}
\usepackage{lipsum}
\usepackage{microtype}

\begin{document}

\title{Computational models in Electroencephalography}

\titlerunning{Computational models in EEG}   

\author{Katharina Glomb \and Joana Cabral \and Anna Cattani \and Alberto Mazzoni \and Ashish Raj \and Benedetta Franceschiello}

\institute{K. Glomb \at
              Department of Radiology, \\ 
              Lausanne University Hospital and University of Lausanne (CHUV-UNIL),\\ 
              Lausanne, Switzerland \\ 
              \email{katharina.glomb@gmail.com} 
           \and
           J. Cabral \at
              Life and Health Sciences Research Institute (ICVS), \\
              University of Minho,\\ 
              Portugal \\
           \and
           A. Cattani \at 
           Department of Psychiatry, University of Wisconsin, \\
           Wisconsin, USA \\
           Department of Biomedical and Clinical Sciences 'Luigi Sacco', University of Milan,\\ Milan, Italy\\
           \and A. Mazzoni \at
           The BioRobotics Institute, \\
           Scuola Superiore Sant'Anna,\\ 
           Pisa, Italy \\
           \and
           A. Raj \at 
           School of Medicine, UCSF, \\ 
           San Francisco, US \\
           \and
           B. Franceschiello \at 
            Department of Ophthalmology, \\ 
            Hopital Ophthalmic Jules Gonin, FAA, \\ 
            Lausanne, Switzerland \\
            Laboratory for Investigative Neurophysiology, \\
            Department of Radiology, \\ 
            Lausanne University Hospital and University of Lausanne (CHUV-UNIL), \\ 
            Lausanne, Switzerland \\ 
            \email{benedetta.franceschiello@gmail.com} 
}

\date{Received: date / Accepted: date} 

\pagestyle{plain}

\maketitle
%
%

\begin{abstract}

Computational models lie at the intersection of basic neuroscience and healthcare applications because they allow researchers to test hypotheses \textit{in silico} and predict the outcome of experiments and interactions that are very hard to test in reality.
Yet, what is meant by ``computational model" is understood in many different ways by researchers in different fields of neuroscience and psychology, hindering communication and collaboration. In this review, we point out the state of the art of computational modeling in Electroencephalography (EEG) and outline how these models can be used to integrate findings from electrophysiology, network-level models, and behavior. On the one hand, computational models serve to investigate the mechanisms that generate brain activity, for example measured with EEG, such as the transient emergence of oscillations at different frequency bands and/or with different spatial topographies. On the other hand, computational models serve to design experiments and test hypotheses \emph{in silico}. The final purpose of computational models of EEG is to obtain a comprehensive understanding of the mechanisms that underlie the EEG signal. This is crucial for an accurate interpretation of EEG measurements that may ultimately serve in the development of novel clinical applications.  

\end{abstract}


\section{Introduction}
\label{intro}
Electroencephalography (EEG) has applications in many fields, spanning from basic neuroscientific research to clinical domains. However, despite the technological advances in recording precision, the full potential of EEG is currently not being exploited. One possible way to do so is to use computational models in order to integrate findings from electrophysiology, network-level models (the level of neuroimaging), and behavior \citep{franceschiello2018neuromathematical, franceschiello2019geometrical}. 
A model is defined in terms of a set of equations which describe the relationships between variables. Importantly, models exist for different spatial scales \citep{varela2001brainweb, deco2008dynamic}, spanning from the single cell spike train up to macroscopic oscillations. The equations are used to simulate how each variable changes over time, or, in rare cases, to find analytical solutions for the relationships among the variables. The dynamics of the resulting time series are also influenced by a set of parameters, which can either be estimated from available data - for example, a model which simulates the firing of a certain neuron type could contain a time constant estimated from recordings on that type of neuron in rodents - or its value can be varied systematically in an exploratory manner. The goal is to produce time series of variables that can be compared to real data. In particular, one can simulate perturbations to brain activity, be it sensory stimulation, a therapeutic intervention like DBS or a drug, or a structural change due to the onset of a pathology, like neurodegeneration or a lesion, and predict what would be the resulting alterations observed in neural and clinical data. 

An important application of EEG models is in the clinical domain. Psychiatric and neurological disorders impact a growing portion of the population, both as patients and caregivers, and with an enormous cost - both economical and humanitarian - to healthcare systems worldwide \citep{steel2014global, vigo2016estimating, feigin2019global}. One of the main obstacles in advancing patient care is the lack of individual diagnosis, prognosis, and treatment planning \citep{wium2017personalized}. Computational models can be adapted to the individual by setting their parameters according to available data (i.e. either setting the parameter directly, if it is measurable, or looking for the parameter value which results in time series whose dynamics match recorded data). The adjusted parameter(s) can then be related to clinical markers, symptoms, and behavior, allowing for example to discriminate between pathologies. Using models in this personalized manner could provide additional diagnostic features in the form of model parameters and model output, eventually assisting clinicians in diagnosis and treatment planning.

Another obstacle is a general lack of scientific knowledge of disease mechanisms, including the mechanisms by which therapies exert their effect. As an example, deep brain stimulation (DBS) is a highly effective treatment for advanced Parkinsonism, in which electric pulses are delivered directly to certain deep brain structures via permanently implanted electrodes. Yet, it is largely unknown how exactly the applied stimulation manages to suppress motor symptoms such as tremor \citep{chiken2016mechanism}. This is also because the way in which motor symptoms result from the degradation of dopaminergic neurons in the substantia nigra is not fully understood \citep{mcgregor2019circuit}. Besides animal models - which have their own ethical issues - \emph{in silico} models are an indispensable tool for understanding brain disorders. Combining data available from a patient or group of patients with knowledge and hypotheses about mechanisms, a model can be generated which can help test these hypotheses. 

Last but not least, models are much cheaper than animal testing or clinical trials. While models will not replace these approaches - at least not in the foreseeable future - they could help to formulate more specific hypotheses and thus, lead to smaller-scale experiments.
Collecting invasive data is not generally possible in humans. EEG is an extremely versatile technology which allows non-invasive recording of neural activity in behaving humans. EEG is a cheap and portable technology, particularly compared to (f)MRI and MEG. Apart from these cost-efficiency considerations, EEG, like MEG, is a direct measure of the electromagnetic fields generated by the brain, and allows millisecond-precision recordings, thus giving access to rich aspects of brain function which can inform models in a way that e.g. fMRI cannot (see section~\ref{sec:eeg} for more details). In general, using different complementary sources of data to construct and validate a model will lead to better model predictions, as each recording technique has its own strengths and weaknesses, and a multimodal approach can balance them. 

In our opinion, there are mostly two reasons why EEG has not been used more extensively in modeling studies, and particularly in a clinical context. First, there are numerous technical problems which make the processing and interpretation of EEG data challenging. EEG - like MEG - is measured on the scalp, and the problem of projecting this 2D-space into the 3D-brain space arises \citep{michel2019eeg}. While multiple solutions exist for this inverse problem, it is unclear which one is the best and under which circumstances \citep{hassan2014eeg, mahjoory2017consistency, hedrich2017comparison}. EEG data require extensive preprocessing, e.g. removal of artifacts due to movements, eye blinks, etc., but these steps are far from being standardized, and many options exist. The recently started EEG-BIDS effort \citep{pernet2019eeg} is a step towards the direction of standardization of EEG data and should facilitate, alongside with the much larger amount of publicly available data, studies that systematically evaluate the impact of preprocessing steps and compare source reconstruction algorithms. As the interest in EEG rises, the need to resolve these issues will trigger larger efforts that will benefit the entire community. 

The second obstacle to a more routine usage of computational models in EEG research, which we hope to address in this review, is that such models usually require an understanding of the mathematics involved, if only to be able to choose the model that is useful for the desired application. 
Both variables and parameters are not always clearly related to quantities which can be measured in a clinical or experimental context, and more generally, models need to be set up in such a way that they meet existing clinical demands or research questions.

The contribution of this paper is threefold. First, this article summarizes computational approaches at different spatial scales in EEG, targeting non-experts readers. To the best of our knowledge, this paper represents the first review on this topic. Second, we will point out several ways in which computational models integrate EEG recordings, by using biologically relevant variables. Third, we discuss the clinical applications of computational models in EEG which have been developed. The field is greatly expanding and contains promising advancements both from research and clinical standpoints. We believe that this overview will make the field accessible for a broad audience, and indicate the next steps required to push modeling of EEG forward. 
\section{Electroencephalography}
\label{sec:eeg}
\begin{figure*}
\centering
    \includegraphics[width = 1\linewidth]{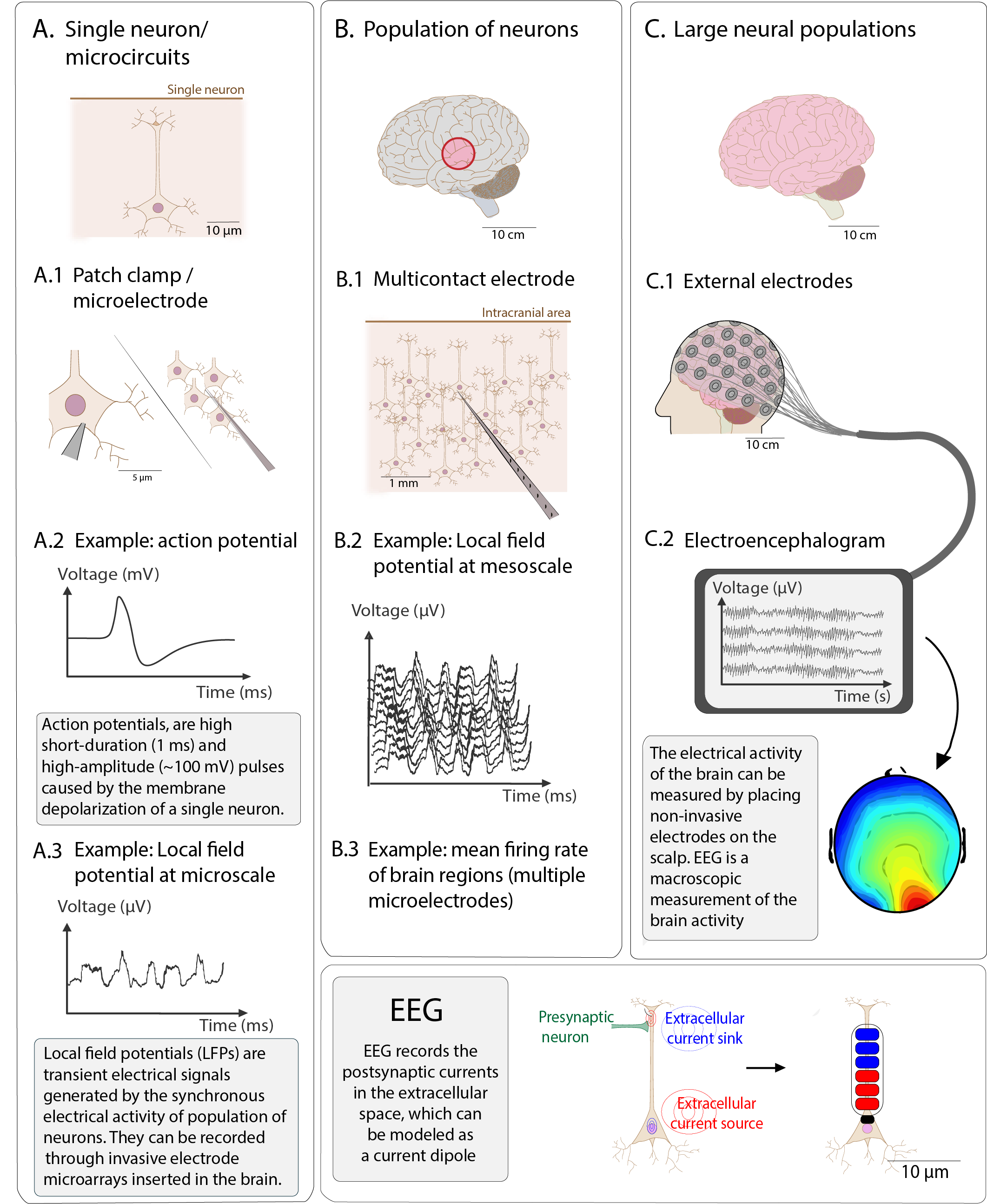}
    \caption{Electrophysiology of neural activity and EEG at different scales.}
     \label{fig:eeg_and_comp_model}
\end{figure*}

EEG is a non-invasive neuroimaging technique that measures the electrical activity of the brain \citep{biasiucci2019electroencephalography}. EEG recordings have been a driver of research and clinical applications in neuroscience and neurology for nearly a century. EEG relies on the placement of electrodes on the person's scalp, measuring the postsynaptic potentials of pyramidal neurons \citep{tivadar2019primer,da2013eeg}. EEG does not directly measure the action potentials of neurons, though there are some indications of high-frequency oscillations being linked to spiking activity \citep{telenczuk2011high}. The neurotransmitter release generated by action potentials, whether excitatory or inhibitory, results in local currents at the apical dendrites that in turn lead to current sources and sinks in the extracellular space around the dendritic arbor (i.e. postsynaptic potentials, see Figure \ref{fig:eeg_and_comp_model}, bottom right block). EEG is generated by the local field potential (LFP), a signal that reflects summed synaptic activity of local populations of neurons. In the neocortex, pyramidal neurons are generally organized perpendicularly to the cortical surface, with apical dendrites toward the pial surface and axons pointing inferiorly towards the grey-white matter border. This alignment leads to the electrical fields of many neurons being summed up to generate a signal that is measurable at the scalp \citep{tivadar2019primer}. Importantly, individual neurons of these populations need to be (nearly) synchronously active to be detectable by EEG.\\  
\indent As mentioned above, the electrical activity of the brain is recorded by means of electrodes, made of conductive materials, placed at the scalp. The propagation of electrical fields takes place due to the conductive properties of brain and head tissues, a phenomenon known as volume conduction \citep{kajikawa2011local}. The electrodes are connected to an amplifier which boosts the signal. Due to the biophysical nature of what is measured, i.e. a voltage - the difference of potential able to move charges from one site to the other - EEG records the differential measurements between an electrode at a specific position on the scalp and a reference site. Common analyses in EEG are the study of local phenomena such as peaks at specific latencies or scalp sites (event-related potential, ERPs); or the study of topography, i.e. the shape of the electric field at the scalp, which represents a global brain signature \citep{murray2008topographic}. EEG is known for its high temporal resolution. The biggest pitfalls of the technique are, on the other hand, the low spatial resolution and signal to noise ratio. A clear and exhaustive walk through these topics as well as an overview of strengths and pitfalls of using EEG is contained in \citet{tivadar2019primer} and for non-experts of the field in \citet{biasiucci2019electroencephalography}.\\
\indent Despite being a measurement of the scalp activity, EEG can reveal the underlying neurophysiological mechanisms of the brain, and that is what classifies it as brain imaging tool. The estimation of the loci of active sources for the recorded brain activity at the scalp is called source reconstruction \citep{michel2004eeg}. However, the loci can belong to areas not necessarily below the considered electrode, a pitfall caused by volume conduction. Source reconstruction is a mathematical ill-posed inverse problem, as the solution is not unique. However, the addition of biophysical constraints to the inverse problem allows to retrieve a solution, which has been validated by means of intracranial recordings \citep{michel2012towards}. Having obtained the source activity, one can estimate the functional connectivity between the sources, i.e. the statistical dependencies between brain areas, assumed to indicate their interactions (see also table~\ref{tab:words}). This can then be complemented with neuroanatomical/structural connectivity (table~\ref{tab:words}), which estimates white matter connections between brain areas. \\ 
\indent Computational models stand at the interface between the physiology of neurons at different scales (single neuron, population, macro-scale) and perceptual behavior. EEG would greatly benefit from the integration of \textit{in-silico} simulations, as computational models could complement both the neurophysiological and behavioral interpretations of EEG recordings. In the following sections, we will discuss different types of computational models, i.e. the different scales at which the neural activity is simulated, how such models can be integrated in the analysis of EEG signals, and how such models have been used in new clinical applications.

\section{Different types of computational models for EEG}
\begin{figure*}
\centering
    \includegraphics[width = 1\linewidth]{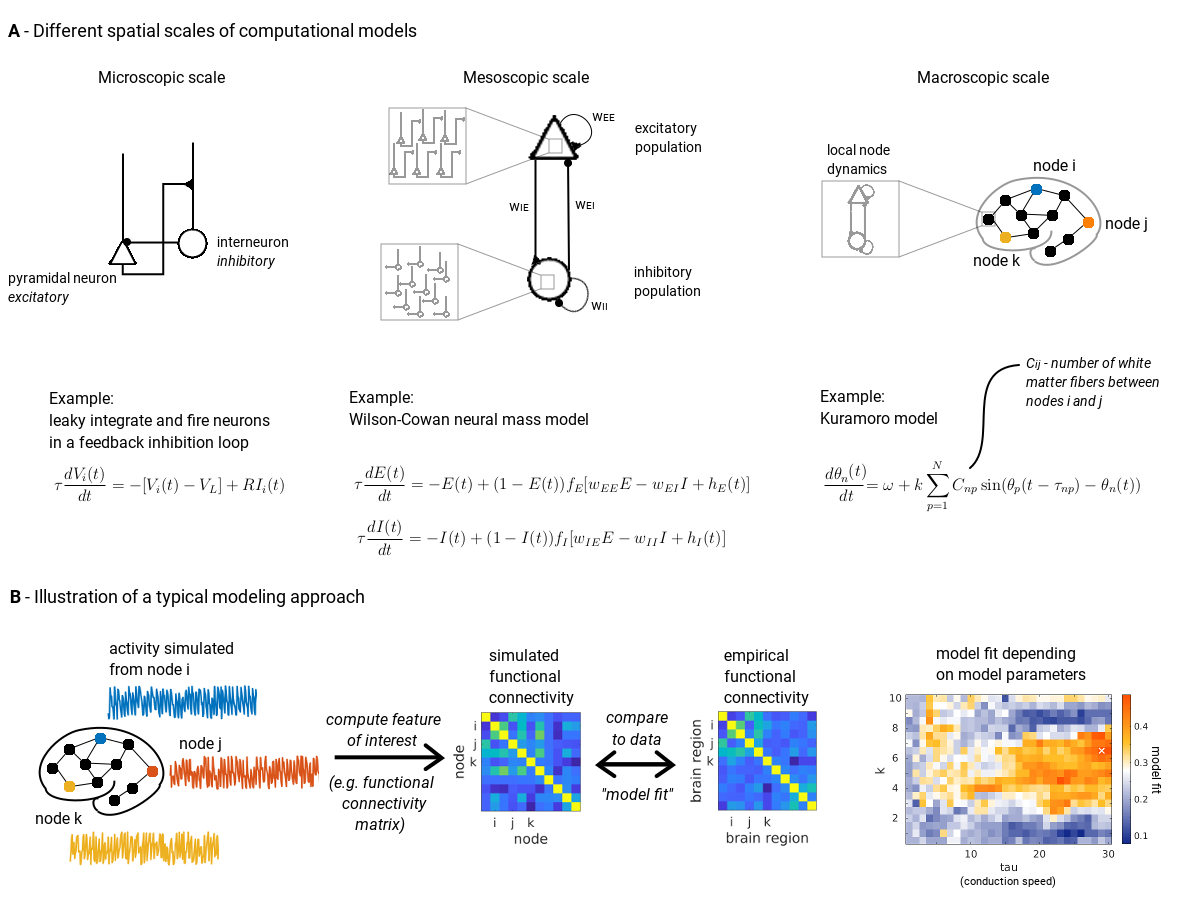}
    \caption{\textbf{A} - Illustration of computational models at the three scales treated here. \emph{Microscopic scale}: Simple example of two ($i=1,2$) leaky integrate-and-fire (LIF) neurons coupled together, a pyramidal neuron making an excitatory synapse to the interneuron, which in turn makes an inhibitory synapse to the pyramidal cell. This minimal circuit implements feedback inhibition, as the pyramidal cell, when activated, will excite the interneuron, which in turn will inhibit it. In the equation, $V_i$ is the membrane potential of each of the two cells $i = 1,2$; $V_L$ is the leak, or resting potential of the cells; $R$ is a constant corresponding to the membrane resistance; $I_i$ is the synaptic input that each cell receives from the other, and possibly background input; $\tau$ is the time constant determining how quickly $V_i$ decays. The model is simulated by setting a firing threshold, at which, when reached, a spike is recorded and $V_i$ is reset to $V_L$. \emph{Mesoscopic scale}: The Wilson-Cowan-model, in which an excitatory ($E$) and an inhibitory ($I$) population are coupled together. The mean field equations describe the mean activity of a large number of neurons. $f_E$ and $f_I$ are sigmoid transfer functions whose values indicate how many neurons in the population reach firing threshold, and $h_E$/$h_I$ are external inputs like background noise. $w_{EE}$ and $w_{II}$ are constants correponding to the strength of self-excitation/inhibition, and $w_{EI}$ and $w_{IE}$ the strength of synaptic coupling between populations. \emph{Macroscopic scale}: In order to simulate long-range interactions between cortical and even subcortical areas, brain network models couple together many mesoscopic (``local") models using the connection weights defined in the empirical structural connectivity matrix $C$. The example equation defines the Kuramoto model, in which the phase $\theta_n$ of each node $n$ is used as a summary of its oscillatory activity around its natural frequency $\omega$. Each node's phase depends on the phases of connected nodes $p$ taking into account the time delay $\tau_{np}$, defined by the distances between nodes $n$ and $p$. $k$ is a global scaling parameter controlling the strength of internode connections. \textbf{B} - Illustration of a typical modeling approach at the macroscopic scale. Activity is simulated for each node using the defined macroscopic model, e.g. the Kuramoto model from panel A, right. The feature of interest is then computed from this activity. Shown here is the functional connectivity, e.g., phase locking values between nodes (table~\ref{tab:words}). This can then be compared to the empirical functional connectivity matrix computed in exactly the same way from experimental data, e.g. by correlating the entries of the matrix. The model fit can be determined depending on parameters of the model, e.g. the scaling parameter $k$ or the unit speed, here indicated with ``tau".}
    \label{fig:model_scales}
\end{figure*}

A straightforward classification of computational models for EEG can be done based on the different scales of neurophysiological activity they integrate. For instance, we can distinguish three types of models (Figure~\ref{fig:model_scales} A):

\begin{enumerate}
    \item microscopic models on the level of single cells and micro-circuits;
    \item mesoscopic models on the level of neural masses and neural fields; 
    \item macroscopic models taking into account the connectome/white matter.  
\end{enumerate}
The integration of computational models has greatly advanced the field of applications of EEG, both for research and clinical purposes. 

\subsection{Computational models for EEG on the level of single cells and microcircuits}

The purpose of this level of modeling is to address the origin of the EEG signal by investigating the relationship between its features and  electrophysiological mechanisms (Figure~\ref{fig:eeg_and_comp_model}, column A) with the tools of computational neuroscience. As detailed above, the EEG signal recorded from the scalp is the result of the spatial integration of the potential fluctuations in the extracellular medium. The EEG signal is mainly caused by the local field potential (LFP), while LFP is mainly driven by synaptic activity \citep{Logothetis3963, Buszaki2012} and volume conduction \citep{kajikawa2011local}. From the experimental standpoint, local network activity is usually measured as LFP (mainly \emph{in vivo} - and rarely \emph{in vitro} - animal data). By virtue of superposition, fluctuations in the LFP, and EEG more generally, are signatures of correlated neural activity \citep{pesaran2018investigating}. Cellular and microcircuit modeling are thus aimed at understanding the neurophysiological underpinnings of these correlations and the role played by cell types, connectivity and other properties in shaping the collective activity of neurons. 

A primary goal of EEG modeling at the microscopic scale is on the one hand to predict the EEG signal generated by the summation of local dynamics on the microscopic scale and, on the other hand, to reconstruct the microscopic neural activity underlying the observed EEG. The first goal is far from being achieved, and the second is ill-posed due to the number of possible circuit and cellular combinations at the source level leading to similar EEGs. Implicit to these goals is to understand how features of neural circuits, such as the architecture, synapses and cell types, contribute to the generation of electromagnetic fields and their properties in a bottom-up fashion. Despite key insights, many shortcomings limit the interpretability of microcircuit models and the establishment of a one-to-one correspondence with EEG data. For instance, the contribution of spiking activity and correlated cellular fluctuations to LFPs and EEG power spectra remains unclear. Most microcircuit models characterize the net local network activity - used as a proxy for EEG - using the average firing rate or via the mean somatic membrane potential taken amongst populations of cells (of various types). Other studies have used a heuristic approach and approximated the EEG signal as a linear combination  of somatic membrane potentials with random coefficients to account for both conduction effects and observational noise \citep{herrmann2016shaping, lefebvre2017stochastic}. As such, microcircuit model predictions and experimental data cannot always be compared directly. 

Cellular multicompartmental models, which oftentimes take cellular morphology and spatial configuration into consideration, are based on the celebrated Hodgkin-Huxley equations, which describe the temporal evolution of ionic flux across neuronal membranes (see \citet{catterall2012hodgkin}, for a recent review). Such conductance-based models, which possess explicit and spatially distributed representations for cellular potentials, facilitate the prediction and/or comparison with LFP recordings. In contrast, single compartment models are difficult to interpret: while more abstract single compartment models such as Poisson neurons or integrate-and-fire models (Figure~\ref{fig:model_scales} A, left) are often used for their relative tractability and computational efficiency to construct more elaborate microcircuit models, they generally lack the neurophysiological richness to estimate EEG traces. Despite this, several computational advancements in recent years investigated how networks of integrate-and-fire neurons generate LFPs, clarifying the microscopic dynamics reflected in the EEG signal \citep{mazzoni2015PCB, mazzoni2008encoding, MAZZONI2010956, MAZZONI20112, deco_dynamic_2008, Buehlmann2008, Barbieri14589}. Such approaches have been used to understand the formation of correlated activity patterns in the hippocampus (e.g. oscillations), and their associated spectral fingerprints in the LFP \citep{chatzikalymniou2018deciphering}. Furthermore, a broad range of works modeled the origin of the local field potential and how it diffuses via volume conduction to generate the EEG signal \citep{hindriks2017linear, linden2011modeling, maki2019biophysical, skaar2019estimation, Gaute2013, telen2017, bedard2009}. 

The key missing element for understanding the link between spiking network activity, LFP, and EEG signal, is the functional and spatial architecture of the networks. In particular, there are two open challenges. The first is to understand how the network connectivity affects the model dynamics that generate the LFP, and the second is to clarify how the spatial arrangement and morphology of neurons affect LFP diffusion \citep{mazzoni2015PCB}.

From this perspective, models of pyramidal cell dynamics and circuits should guide the interpretation of the EEG signal. For example, Destexhe and colleagues recently addressed the long-debated issue of the relative contribution of inhibitory and excitatory signals to the extracellular signal \citep{telenczuk2019modeling}, suggesting that the main source of the EEG signal may stem from inhibitory - rather than excitatory - inputs to pyramidal cells. A recent spiking network model \citep{saponati2019integrate} incorporates the modular architecture of the thalamus, in which subnetworks connect to different parts of the cortex \citep{barardi2016}. This model was used to show how the propagation of activity from the thalamus shapes gamma oscillations in the cortex. 

Computational models at the level of single cells and microcircuits have also been instrumental in elucidating the mechanisms underlying multiple EEG phenomena. For instance, such models  were used to better understand EEG rhythm changes observed before, during and after anesthesia, using spiking network models \citep{Kopell2008, Kopell2012} and/or cortical micro-circuit models \citep{hutt2018suppression}. Some of these models have been extended to account for the effect of thalamocortical dynamics on EEG oscillations \citep{ching2010thalamocortical, hutt2018suppression}, highlighting the key role played by the thalamus on shaping  EEG dynamics. In addition, microcircuit models have been used to understand the EEG response of cortical networks to non-invasive brain stimulation (e.g. TACS, TMS), especially in regard to the interaction between endogenous EEG oscillatory activity and stimulation patterns \citep{herrmann2016shaping}, in which thalamic interactions were found to play an important role \citep{lefebvre2017stochastic}. 

\subsection{Computational models for EEG on the level of neural masses and neural fields} \label{subsec:neural_masses}

In this section we discuss models of population dynamics and how they could determine specific features of the electrical activity recorded by EEG (Figure~\ref{fig:eeg_and_comp_model}, column B). Mean field models describe the average activity of a large population of neurons by modeling how the population - as a whole - transforms its input currents into an average output firing rate (Figure~\ref{fig:model_scales} A, middle; for details on how networks of spiking neurons are reduced to mean field formulations, see \citep{wong2006recurrent,deco2013resting,coombes2019next,byrne2020next}). If we consider a population to be a small portion of the cortex containing pyramidal cells, the average activity modelled by the mean field can be understood as the LFP. Two types of models can be distinguished: neural mass models, where variables are a function of time only, and neural field models, where variables are functions of time and space. In this sense, neural field models can be seen as an extension of neural mass models, by taking into account the continuous shape of cortical tissue and the spatial distribution of neurons. These models allow for the description of local lateral inhibition as well as local axonal delays \citep{hutt2003pattern,atay2006neural}. An important application of neural field models is found in phenomenological models of visual hallucinations \citep{ermentrout1979mathematical,bressloff2001geometric}, and they have been used to model sleep and anaesthesia \citep{steyn1999theoretical,bojak2005modeling}. Future applications may also involve both neural mass and neural field models to describe different cortical structures, similarly to the multiscale approach proposed in \citet{cattani2016hybrid}.

The most popular model on this mesoscopic scale was first described by Wilson \& Cowan \citep{wilson1973mathematical,cowan2016wilson} (Figure~\ref{fig:model_scales} A, middle), and all mean field models can be seen as deriving from this form. It consists of an inhibitory and an excitatory population, where usually, for the purpose of EEG, it is assumed that the excitatory population models pyramidal neurons while the inhibitory population takes the role of interneurons. A variant of this model was described in Jansen \& Rit \citep{jansen1995electroencephalogram} and goes back to the ``lumped parameter" model by Lopes da Silva \citep{da1974model}. It uses three distinct populations, i.e. a population of excitatory interneurons in addition to the two populations already mentioned. The reason this model has been popular in EEG modeling is that it accounts for the observation that inhibitory and excitatory synapses tend to deliver inputs to different parts of the pyramidal cell body \citep{sotero2007realistically}. In addition, thalamocortical loops are thought to greatly contribute to the generation of oscillations observed in the cortex \citep{steriade1993thalamocortical}, and an important class of neural field models deals with these loops and their dependency on external stimuli \citep{robinson2001prediction, robinson_dynamics_2002}.

The dynamical behavior of models can be manipulated to simulate different phenomena by varying their parameters. For example, the coupling parameters that determine the strength and speed of feedback-inhibition and feedforward-excitation can be varied (parameters $w_{IE}$ and $w_{EI}$ in Figure~\ref{fig:model_scales} A, middle), both within and between populations. Also it is possible to modify time constants (which govern the decay of activity in the local populations) or the strength of background noise. Changing these parameters \emph{in silico} can be interpreted biologically. For example, in \citet{bojak2005modeling}, the authors describe how a modified neural field model reproduces EEG spectra recorded during anaesthesia. The strength of inputs from the thalamus to the cortical neural populations was varied within a biologically plausible range.

Neural mass and neural field models are able to reproduce a range of dynamical behaviors that are observed in EEG, like oscillations in typical EEG frequency bands \citep{david2003neural}, phase-amplitude-coupling \citep{onslow2014canonical,sotero2016topology}, evoked responses \citep{jansen1993neurophysiologically,jansen1995electroencephalogram,david2005modelling}, and allows to model the EEG spectrum \citep{david2003neural,bojak2005modeling,moran2007neural}. 

By coupling together more than one model/set of populations, one can start investigating the effect that delays have on neural activity \citep{jirsa1996field}. In fact, Jansen \& Rit \citep{jansen1995electroencephalogram}  coupled together two neural mass models in order to simulate the effect of interactions between cortical columns on their activity. 

Often, activity simulated by mean field models is assumed to be related to local field potentials \citep{liley2002spatially}. However, models are usually set up such that the local field potential derives directly from the mean firing rate. In this way, an important aspect that underlies the EEG signal is neglected, namely, the synchrony (coherence) of the firing within a neural population (as opposed to synchrony between populations, which can be studied using e.g. instantaneous phase differences \citep{breakspear2004novel}). Phenomena such as event-related synchronization and -desynchronization result from a change in this synchrony rather than from a change in firing rate. Recent models \citep{byrne2017mean,byrne2020next} propose therefore a link between the firing rate and the Kuramoto order parameter, which is a measure of how dispersed firing is within a population. 

\subsection{Macroscopic computational models for EEG taking into account the connectome}\label{subsec:macro}


In this section, we review existing literature on macroscopic computational models that take into account the connectome and discuss their potential to reveal the generative mechanisms of the macroscopic brain activity patterns detected with EEG and MEG (Figure~\ref{fig:eeg_and_comp_model}, column C). We will use the term ``brain network models" (BNM) in order to clearly distinguish this framework from other approaches to whole-brain modeling \citep{breakspear2017dynamic}, e.g. using neural field models \citep{jirsa1996field,robinson1997propagation,coombes2007modeling} or expansions of the thalamocortical models discussed above \citep{robinson2001prediction,freyer2011biophysical}. We will also leave aside the large body of literature on dynamic causal modeling (DCM) \citep{kiebel2008dynamic,pinotsis2012dynamic}, as this deserves a more detailed review than the scope of this paper can provide. 

\paragraph{Brain network models.} In recent years, the interest in the human connectome has experienced a boom, creating the prolific and successful field of ``connectomics". In the framework of connectomics, the brain is conceptualized as a network made up of nodes and edges. Each node represents a brain region, and nodes are coupled together according to a weighted matrix representing the wiring structure of the brain (Figure~\ref{fig:model_scales} A, right). This so-called structural connectivity matrix (SC) is derived from white matter fiber bundles which connect distant brain regions \citep{behrens2003non,zhang2010noninvasive,hagmann2008mapping,sepulcre2010organization,wedeen2012geometric} and are measured using diffusion weighted magnetic resonance imaging (dMRI) (table~\ref{tab:words}). The set of all fiber bundles is called the connectome \citep{sporns2011human}. By coupling brain regions together according to the weights in the SC, the activity generated in each region depends also on the activity propagated from other regions along the connections given by the SC. 

BNMs are used to study the role of structural connectivity in shaping brain activity patterns. Because this is a complex problem that involves the entire brain, it is important to find a balance between realism and reduction, so that useful predictions can be made.
In practice, a common simplification is to assume that all brain regions are largely identical in their dynamical properties \citep{passingham2002anatomical}. This \emph{reductionist} approach keeps the number of parameters at a manageable level and still allows to investigate how collective phenomena emerge from the \emph{realistic} connectivity between nodes. 
In other words, BNMs do not necessarily aim at maximizing the fit to the empirically recorded brain signals. Rather, the goal is to reproduce specific temporal, spatial or spectral features of the empirical data emerging at the macroscopic scale whose underlying mechanisms remain unclear (Figure~\ref{fig:model_scales} B).

\paragraph{Choice of local model.} In mathematical terms, brain activity is simulated according to a system of coupled differential equations. The activity of each node is described by a mean-field model, such as the ones described in section \ref{subsec:neural_masses}, and coupling between the mean field models is parametrized by the empirical SC (Figure~\ref{fig:eeg_and_comp_model} A, right). 

Importantly, the type of mean-field model used at the local level must be selected according to the hypothesis being tested. For example, BNMs have proved to be a powerful tool to elucidate the non-linear link between the brain's structural wiring and the functional patterns of brain activity captured with resting-state functional magnetic resonance imaging (rsfMRI) \citep{deco_identification_2014, deco_resting_2013, honey_predicting_2009, deco2009key, cabral2011role}. However, oscillations in frequency ranges important for M/EEG (2-100 Hz) are often neglected in studies aiming at reproducing correlated fluctuations on the slow time scale of the fMRI signal. Thus, despite the insights gained by BNMs to understand rsfMRI signal dynamics, the same models do not necessarily serve to understand M/EEG signals and vice-versa \citep{cabral2017functional}.

In \citet{cabral2014exploring}, the local model employed includes a mechanism for the generation of collective oscillatory signals in order to address oscillatory components of M/EEG. To model brain-wide interactions  between local nodes oscillating around a given natural frequency (in this case, 40 Hz, in the gamma frequency range), the Kuramoto model \citep{kuramoto2003chemical,yeung1999time}, was extended to incorporate realistic brain connectivity (SC) and time delays (determined by the lengths of the fibers in the SC (see also \citet{finger2016modeling}; Figure~\ref{fig:model_scales} A, right). This model shows how, for a specific range of parameters, groups of nodes (communities) can temporarily synchronize at community-specific lower frequencies, obeying to universal rules that govern the behaviour of coupled oscillators with time delays. Thus, the model proposed a mechanism that explains how slow global rhythms in the alpha- and beta-range emerge from interactions of fast local (gamma) oscillations generated by neuronal networks. 

In contrast, \citet{deco2019brain} used a mean field model \citep{wilson1973mathematical, brunel2001effects, deco2012ongoing, deco2014local}, which was tuned not to exhibit intrinsic oscillations. Because the brain could thus be considered as being in a noisy, low-activity state, the number of parameters was sufficiently reduced to investigate how activation patterns change over time on different time scales. Time scales including that of M/EEG (ten to several 100 ms) as well as that typical for fMRI (1-3 seconds) were considered, and the question was asked whether there is a time scale at which brain dynamics are particularly rich. The authors found that the best frequency resolution was on the scale of 200 ms, as both the number of co-activation patterns as well as their dynamics were richest, compared to other resolutions. 

\paragraph{Emerging class of harmonics-based models.}
Although both the described BNMs as well as DCM (dynamic causal modeling) have a long history of success in modeling brain activity patterns, they have high-dimensionality, and usually require local oscillators governed by region-specific or spatially-varying model parameters. While this imbues such models with rich features capable of recreating complex behavior, they are challenging for some clinical applications where a small set of global features might be desired to assess the effect of disease on network activity. Therefore recently some laboratories have focused on low-dimensional processes involving diffusion or random walks  (table~\ref{tab:words}) on the structural graph  (table~\ref{tab:words}) instead of mean-field models, providing a simpler means of simulating functional connectivity (FC). These simpler models were able to match or exceed the predictive power of complex neural mass models or DCMs in predicting empirical FC \citep{abdelnour2014network}. Higher-order walks on graphs have also been quite successful; typically these methods involve a series expansion of the graph adjacency or Laplacian matrices \citep{Meier2016, Becker2018}  (table~\ref{tab:words}). Not surprisingly, the diffusion and series expansion methods are closely related, and most of these approaches may be interpreted as special cases of each other, as demonstrated elegantly in recent studies \citep{robinson2016eigenmodes, deslauriers2020, tewarie2020}.

Whether using graph diffusion or series expansion, these models of spread naturally employ the so-called eigenmodes, or harmonics, of graph adjacency or Laplacian matrix. Hence these methods were generalized to yield spectral graph models whereby e.g. Laplacian harmonics were sufficient to reproduce empirical FC, using only a few eigenmodes \citep{Atasoy2016, Abdelnour2018}. The Laplacian matrix in particular has a long history in graph modeling, and its eigenmodes are the orthonormal basis of the network and can thus represent arbitrary patterns on the network \citep{Stewart1999}. Such spectral graph models are computationally attractive due to low-dimensionality and more interpretable analytical solutions. The SC's Laplacian eigenmodes may be thought of as the substrate on which functional patterns of the brain are established via a process of network transmission \citep{Abdelnour2018, Atasoy2016, robinson2016eigenmodes, preti_decoupling_2019, glomb2020connectome}. These models were strikingly successful in replicating canonical functional networks, which are stable large scale circuits made up of functionally distinct brain regions distributed across the cortex that were extracted by clustering a large fMRI dataset \citep{Yeo2011}. 

While spectral graph models have demonstrated ability to capture essential steady-state, stationary characteristics of real brain activity, they are limited to modeling passive spread without oscillatory behavior. Hence they may not suitably accommodate a larger repertoire of dynamically-varying microstates or rich power spectra at higher frequencies typically observed on EEG or MEG. Capturing this rich repertoire would require a full accounting of axonal propagation delays as well as local neural population dynamics within graph models, as previously advocated \citep{cabral2011role}. Band-specific MEG resting-state networks were successfully modeled with a combination of delayed neural mass models and eigenmodes of the structural network \citep{tewarie2019spatially}, suggesting delayed interactions in a brain's network give rise to functional patterns constrained by structural eigenmodes. Recently another effort was undertaken to characterise wide-band brain activity using graph harmonics in closed form (i.e. requiring no time-domain simulations), a rarity in the field of computational neuroscience  \citep{Raj2020}. This  ``spectral graph model" of brain activity produced realistic power spectra that could successfully predict both the spatial as well as temporal properties of MEG empirical recordings \citep{Raj2020}. Intriguingly, the model has very few (six) parameters, all of which are global and not dependent on local oscillations. This method therefore exemplifies the power of graph methods in reproducing more complex and rich repertoire of brain activity, while keeping to a parsimonious approach that does not require the kinds of high-dimensional and non-linear oscillatory models that have traditionally held sway.

\section{Applications of computational models of EEG}

Network oscillations, captured through EEG, are thought to be relevant for brain functions, such as cognition, memory, perception and consciousness \citep{ward_synchronous_2003}. Local brain regions produce oscillatory activity that propagates through the network to other brain regions. Alterations of oscillatory activity can be a sign of a brain disorder, and they are thought to be due to changes at the level of tissue and local/global connectivity. Due to its ability to capture such oscillatory activity, EEG is commonly used in research and clinical fields to study the neurophysiological bases of brain disorders, helping diagnosis and treatment \citep{iv_handbook_2014}. Physiologically and theoretically inspired computational models are able to reproduce EEG signals, offering a unique tool which complements experimental approaches. The application of computational models reveals disease mechanisms, helps testing new clinical hypotheses, and to explore new surgical strategies \emph{in silico}. This section presents computational models of EEG that have been employed to study different states of consciousness - wakefulness, deep sleep, anesthesia, and disorders of consciousness - as well as diseases such as neuropsychiatric disorders and epilepsy.

\subsection{States of consciousness}\label{subsec:DoC}

A variety of models have been employed to investigate the brain dynamics in different physiological brain states, such as wakefulness and deep sleep (non-rapid eye movement, NREM) \citep{hill_modeling_2005,cona_thalamo-cortical_2014,robinson_dynamics_2002,roberts_corticothalamic_2012}, and pharmacological conditions, such as anesthesia \citep{ching_modeling_2014,ching_thalamocortical_2010,sheeba_neuronal_2008,hutt_effects_2010,liley_propofol_2010}. Other modeling approaches seek to elucidate the neurophysiological mechanisms underlying the presence or the absence of consciousness in wakefulness, NREM sleep and anesthesia, and they have crucial implications for the study of disorders of consciousness (DoC). DoC refer to a class of clinical conditions that may follow a severe brain injury (hypoxic/ischemic or traumatic brain injury) and include coma, vegetative state or unresponsive wakefulness syndrome (VS/UWS), and minimally conscious state (MCS). Coma has been defined as a state of unresponsiveness characterized by the absence of arousal (patients lie with their eyes closed) and, hence, of awareness. VS/UWS denotes a condition of wakefulness with reflex movements and without behavioural signs of awareness, while patients in MCS show unequivocal signs of interaction with the environment. 

The current gold standard for clinical assessment of consciousness relies on the Coma Recovery Scale Revised \citep{giacino2004jfk}, which scores the ability of patients to behaviourally respond to sensory stimuli or commands. However, behavioral-based clinical diagnoses can lead to misclassification of MCS as VS/UWS because some patients may regain consciousness without recovering their ability to understand, move and communicate \citep{childs_accuracy_1993,andrews_misdiagnosis_1996,schnakers_diagnostic_2009}. A great effort has been devoted to develop advanced imaging and neurophysiological techniques for assessing covert consciousness and to improve diagnostic and prognostic accuracy \citep{edlow_early_2017,bodart_measures_2017,stender_diagnostic_2014,bruno_unresponsive_2011,owen_detecting_2008,stender_minimal_2016}. A novel neurophysiological approach to unravel the capacity of the brain to sustain consciousness exploits Transcranial Magnetic Stimulation (TMS) in combination with EEG \citep{rosanova_sleep-like_2018, casarotto_stratification_2016}. Specifically, the EEG response evoked by TMS in \emph{conscious} subjects exhibits complex patterns of activation resulting from preserved cortical interactions. In contrast, when \emph{unconscious} patients are stimulated with TMS, the evoked-response shows a local pattern of activation, similar to the one observed in healthy controls during NREM sleep and anesthesia. 

The perturbational complexity index (PCI) \citep{casali_theoretically_2013} captures the dynamical complexity of TMS-evoked EEG potentials by means of the Lempel-Ziv compression algorithm, showing high values (low compressibility) for complex chains of activation typical of the awake state, and low values (high compressibility) for stereotypical patterns of activation typical of sleep and anesthesia. PCI has been validated on a benchmark population of 150 conscious and unconscious controls and tested on 81 severely brain-injured patients \citep{casarotto_stratification_2016}, showing an unprecedented high sensitivity (94.7\%) in discriminating conscious from unconscious states.

A recently published modeling approach \citep{bensaid_coalia_2019} investigates the physiological mechanisms underlying the generation of complex or stereotypical TMS-evoked EEG responses. The proposed brain network model, named COALIA, describes local dynamics as neural masses \citep{wendling_epileptic_2002} that include populations of pyramidal neurons and three different types of interneurons. Each neural mass describes the local field activity of one of 66 cortical brain regions \citep{desikan_automated_2006}. Neural masses are connected with each other through long-range white matter fibers as described above (section~\ref{subsec:macro}). EEG signals are then simulated as neural mass activity. A systematic comparison of the complexity of simulated and real TMS-evoked EEG potentials through PCI suggested that the rhythmically patterned thalamocortical activity, typical of sleep, plays a key role in disrupting the complex patterns of activation evoked by TMS \citep{bensaid_coalia_2019}. Indeed, this rhythmical thalamocortical activity results in inhibition within the cortex that prevents information from propagating from one brain region to another, and thus disrupts functional integration, i.e. the ability of the brain to integrate information originating from different brain regions or groups of brain regions \citep{tononi_consciousness_1998}. Functional integration is necessary, along with functional segregation, i.e. the ability of brain regions or groups of brain regions to fulfill a certain function distinct from other areas of the brain \citep{lord_understanding_2017}, to generate complex time-varying patterns of coordinated cortical activity that are typical of the awake brain, and thought to sustain consciousness and cognitive functions \citep{casali_theoretically_2013,demertzi_human_2019}.

\subsection{Neuropsychiatric disorders}

Disruption of integration and segregation balance, which is fundamental for consciousness as mentioned in section~\ref{subsec:DoC}, have been linked
also to several neuropsychiatric disorders as a result of altered structural and functional connectivity \citep{bassett_human_2009,fornito_connectomics_2015, menon_large-scale_2011, deco_rethinking_2015}. Among neuropsychiatric disorders, as reviewed in \citet{lord_understanding_2017}, Alzheimer's disease is characterized by a decrease in long-range functional connectivity, directly affecting integration between functional modules of the brain \citep{stam_small-world_2007,sanz-arigita_loss_2010}. Schizophrenia has been linked to a ``subtle randomization" of global functional connectivity, such that the so-called ``small-world" character of the network is disrupted \citep{alexander-bloch_disrupted_2010,lynall_functional_2010}; a small-world network is characterized by short path lengths and strong modularity, network properties that are thought to promote information processing in the brain \citep{bassett2006small} (but see \citet{hilgetag2016brain}). Loss of integration has also been observed in schizophrenia \citep{damaraju_dynamic_2014}.

As explained in section~\ref{subsec:macro}, whole-brain computational models provide insights into how anatomical connections shape and constrain functional connectivity \citep{deco_identification_2014,deco_resting_2013,honey_predicting_2009}. Using BNMs, Cabral and colleagues have shown that the alterations reported in schizophrenia \citep{lynall_functional_2010} can be explained by a decrease in connectivity between brain areas, occurring either at the local or global level and encompassing either axonal or synaptic mechanisms, hence reinforcing the idea of schizophrenia being the behavioural consequence of a multitude of causes disrupting connectivity between brain areas \citep{cabral2012functional, cabral2012modeling}.

However, these models have focused on reproducing fMRI findings and are yet to be extended to address alterations in EEG spectral signatures in schizophrenia, namely increased EEG gamma-band power and decreased alpha power \citep{uhlhaas2013high}, which, following previous modeling insights \citep{cabral2014exploring}, may arise from reduced coupling between local gamma-band oscillators. Furthermore, BNMs can be employed to test how clinical interventions may help to re-establish healthy network properties such as balance between integration and segregation or small-worldness \citep{deco2018perturbation, deco2019awakening}. 

\subsection{Epilepsy}\label{sec:epilepsy}

Models have been employed to study pathological alterations of network oscillatory activity related to many diseases, including epilepsy \citep{stefanescu_computational_2012, wendling_neurocomputational_2005, lytton_computer_2008, holt_computational_2013}. Epilepsy is a complex disease which impacts 1\% of the world population and is drug resistant in approximately 30\% of cases. Due to its intrinsic complexity, epilepsy research  has strongly benefited, and will do so even more in the future, from an \emph{in silico} environment where hypotheses about brain mechanisms of epileptic seizures can be tested in order to guide strategies for surgical, pharmacological and electrical stimulation techniques.

Focal epilepsy is a prototypical example of a disease that involves both local tissue and network properties. Focal epilepsy occurs when seizures originate in one or multiple sites, so-called epileptogenic zones (EZ), before recruiting close and distal non-epileptogenic areas pertaining to the pathological network. Patients with a history of drug-resistant focal epilepsy are candidates for surgery which targets epileptogenic areas and/or critical nodes presumably involved in the epileptogenic network. Successful outcomes of these procedures critically rely on the ability of clinicians to precisely identify the EZ. 

A promising modeling approach aims at studying focal epilepsy through a single-subject virtual brain \citep{proix_individual_2017, terry_seizure_2012, hutchings_predicting_2015, bansal_personalized_2018, soltesz_computational_2011}, bringing together the description of how seizures start and end (seizure onset and offset, respectively) at a local level (through neural mass models) \citep{robinson_dynamics_2002, wendling_epileptic_2002, lopes_da_silva_epilepsies_2003, breakspear_unifying_2006, jirsa_nature_2014} with individual brain connectivity derived from dMRI data. In this personalized approach, a patient's brain is virtually reconstructed, such that systematic testing of many surgical scenarios is possible. The individual virtual brain approach provides clinicians with additional information, helping them to identify locations which are responsible for starting or propagating the seizure and whose removal would therefore lead to the patient being seizure-free while avoiding functional side effects of removing brain regions and connections \citep{olmi_controlling_2019,proix_individual_2017}. 

\begin{table*}
    \caption{Some terminology used in this paper.}
    \centering
    \begin{tabular}{p{0.25\textwidth}|p{0.7\textwidth}}
         Functional connectivity &  (FC) Statistical dependencies between time series recorded from different brain regions or simulated at different nodes. Such dependencies are taken to indicate a functional relatedness of the brain regions/nodes. Many measures are available, for example correlation between amplitude envelopes, phase locking value, imaginary coherence, etc. See for example \citet{colclough2016reliable} for an overview. Note that FC does not establish a causal relationship \citep{friston2011functional}. \\
         \\
         Structural connectivity & (SC) Also known as neuroanatomical, anatomical, or white matter connectivity. Diffusion-weighted MRI (dMRI) is able to measure the diffusion of water through brain tissue \citep{basser1994mr}. As water diffuses preferably \emph{along} axons rather than across their walls, the orientation of large fiber bundles can be inferred from dMRI via algorithms known as fiber tracking \citep{jones2010challenges}. Note that SC does not take into account local anatomical connections made within the gray matter, and that fiber counts or densities do not allow making conclusions about the weight of that connection \citep{jeurissen2019diffusion}. Furthermore, fiber tracking algorithms are unable to resolve ambiguities introduces by crossing fibers, and it is difficult to track long fibers. \\
         \\
         Graph & A brain network model, which consists of nodes and edges (Figure~\ref{fig:model_scales} A, right), can be formalized as a graph \citep{bassett2017network, sporns2018graph}. This can be visualized using so-called adjacency matrices, which contain a weight for the edge between each pair of nodes ((Figure~\ref{fig:model_scales} B)). In this sense, both FC and SC matrices are adjacency matrices. This formalization opens up the analysis of brain networks to the tools of graph analysis. These tools allow for example the characterization of the graph/network using many different quantitative measures \citep{rubinov2010complex}, partitioning the graph/network into subnetworks or modules \citep{bassett2017network, donetti2004detecting}, or classifying nodes depending on their role in the network \citep{hagmann2008mapping}.\\ 
         Random walk & A random walk is a random process taking place on the graph in which a ``walker" is initiated at a node and proceeds to another node following existing edges. Edges are selected by the walker with a probability proportional to their weight. Such a simulation can be used to approximate the dynamics of spreading activation, and enables the researcher to approximate for example the probability that activity will spread from node $i$ to node $j$ given the edges that exist between them, or the time that it will take for activity to spread from node $i$ to node $j$.  \\
         \\
         Laplacian & The Laplace operator is ubiquitous in many physical systems and is used to describe standing waves, diffusion, heat dispersion, and many other phenomena. For a network, the Laplacian is obtained directly from the adjacency matrix (see above). An intuitive interpretation is that it describes the ``flow" of activity along the edges.  
         \\
         Eigenmodes & Many physical systems that consist of interacting elements can vibrate at certain frequencies, for example the string of a violin or the vibrating sheets of Chladni \citep{chladni1802akustik}. Each system has its own set of frequencies at which it can vibrate, determined for example by its shape. Mathematically, these eigenmodes are obtained via eigendecomposition of the Laplacian (see above). \\
    \end{tabular}
    \label{tab:words}
\end{table*}

\section{Discussion}


In this paper we introduced different computational model types and their application to EEG, using a simple classification by spatial scale. Clearly not all models in the literature would necessarily belong to one category, but we believe this taxonomy can provide an entry point for non-experts. The main motivation behind this review was to identify obstacles that stand in the way of applying EEG modeling in both a research and clinical context, and to point out future directions that could remove these obstacles. 

We have pointed out several recent efforts that have begun to more closely align models and experimental findings. Such integration of theory and experiment guarantees the use of biologically relevant measures within computational models of EEG, a crucial element if one wishes to use EEG models together with acquired data. For example, recent microcircuit models address the gap between theory and experiment by linking average firing rate - a measure of population activity preferred by the modeling community - and local field potential (LFP) - a measure that is generally thought to be a good proxy of the EEG signal \citep{saponati2019integrate}; recent mean field models explicitly include the contribution of neural synchronization to the LFP \citep{byrne2020next}, thereby integrating experimental knowledge about how the EEG signal is generated \citep{da2013eeg}; brain network models explore the contribution of empirically measured connectomes to macroscopic brain activity \citep{cabral2014exploring}; and applications of computational models already exist that use clinical measures to study e.g. coma \citep{bensaid_coalia_2019}, epilepsy \citep{olmi_controlling_2019,proix_individual_2017}, and neuropsychiatric disorders \citep{spiegler_selective_2016, kunze_transcranial_2016}. Furthermore, some modeling approaches focus on providing a simple model for large-scale dynamics, making results more interpretable both from a theoretical and clinical standpoint \citep{Abdelnour2018,Raj2020}. 

We have reviewed computational models on three spatial scales (Figure~\ref{fig:model_scales} A). Each scale models qualitatively different biological processes which can be measured using distinct recording techniques \citep{varela2001brainweb} (Figure~\ref{fig:eeg_and_comp_model}). While EEG records activity at the macro-scale, mechanisms at each scale have an impact on the EEG signal and should therefore inform its interpretation. Therefore, ideally, scales should be combined to provide a complete picture of neural mechanisms underlying EEG activity, something that started to be explored for example in the simulation platform The Virtual Brain (TVB) \citep{sanz2013virtual, falcon2016new} or in studies showing the theoretical relationship between spiking networks and mean field formulations \citep{wong2006recurrent, deco2013resting, coombes2019next, byrne2020next}. Using models in this hierarchical manner is the only way to disentangle different contributions to the EEG signal without using invasive techniques, i.e., to distinguish neural signals \citep{michel2012towards, seeber2019subcortical}, volume conduction \citep{hindriks2017linear, linden2011modeling, maki2019biophysical, skaar2019estimation, Gaute2013, telen2017, bedard2009}, and noise. Furthermore, brain disorders can impact brain structure and function on any scale. Using models on multiple scales is necessary if one wishes to understand how pathological changes manifest in clinically measurable EEG signals. Such an understanding would also allow to use EEG to evaluate clinical interventions that affect the micro- or mesoscale (e.g., drugs).

Models can thus play an important role as a ``bridge" that connects different fields. In translational applications, knowledge from basic research can be integrated into a model and the model can be designed in such a way that it is useful for a clinical application. An example for a successful ``bridge" is the case of Brain Computer Interfaces. In order to realize multi-scale models, researchers working on animal recordings and researchers focusing on non-invasive recordings in humans have to come together with modeling experts that can incorporate findings from both fields in a model. 

As an outlook, EEG modeling could play an important part in future endeavors towards precision medicine, or ``personal health". Individual brain models could be used to integrate different sources of data (EEG, fMRI, ECG, etc.) in a ``virtual patient". This could complement data-driven approaches like connectome fingerprinting (in which individuals are identified using their individual connectome \citep{finn2015functional, pallares2018extracting, abbas2020geff}). The ultimate goal would be to use this virtual patient to tailor diagnosis and therapies around the needs of the patient \citep{wium2017personalized}, reducing the economical burden and patient discomfort of clinical analyses and hospitalisation. 

\begin{acknowledgements}
The authors would like to thank Ana Hernando Ariza for the first image contained in this paper and the creativity, time and efforts she put into transforming the important messages of this article in graphics. The authors are particularly grateful to Prof. J\'{e}r\'{e}mie Lefebvre and Prof. Micah M. Murray for their contributions, guidance and mentoring during the preparation of this article.
\end{acknowledgements}

\section*{Funding}
K.G. was funded by Schweizerischer Nationalfonds zur F\"{o}rderung der Wissenschaftlichen Forschung Award ID: 170873. J.C. was funded by the Portuguese Foundation for Science and Technology (FCT) CEECIND/03325/2017 and by projects UID/Multi/50026 (FCT), NORTE-01-0145-FEDER- 000013 and NORTE-01-0145-FEDER-000023. A.C. was supported by the Tiny Blue Dot Foundation. A.M. was supported by PREVIEW - Bando salute 2018 - Tuscany Regional Government. A.R. was supported by NIH grants R01NS092802, RF1AG062196 and R56AG064873. B.F. received financial support for this work by the Fondation Asile des aveugles (grant number 232933), a grantor advised by Carigest SA (number 232920). 

\section*{Conflict of interest}
The authors declare that they have no conflict of interest.

\section*{Authors contribution}
K.G. and B.F. conceived the review. J.C. and A.R. contributed to section 2.2. A.M. contributed to section 2.1. A.C. entirely conceived section 3. K.G. and B.F. outlined all remaining contents and all authors contributed to the final draft and review. 
%

\bibliographystyle{spbasic}      
\bibliography{bibliography}   

%
%

\end{document}